\documentclass{mem}
\usepackage{natbib}\usepackage{txfonts}%\usepackage{balance}
\usepackage{graphicx}
\usepackage[a4paper]{hyperref}
\idline{75}{282}
\begin{document}

%   \title{Oscillation mode lifetimes in $\xi\,$Hydrae: Will strong mode damping limit asteroseismology of red giant stars?} %PAPER TITLE
%   \title{Strong mode damping can kill asteroseismology: Measuring the mode lifetime of the red giant $\xi\,$Hydrae} %ORG ABSTRACT TITLE
%   \title{Are mode lifetimes of red giant stars too short?} %TALK TITLE
   \title{Do red giants have short mode lifetimes?}

\titlerunning{Are mode lifetimes of red giant stars too short?}

   \author{D. Stello\inst{1,2,3}
          \and
          H. Kjeldsen\inst{1}
          \and
          T. R. Bedding\inst{2}
          \and
          D. Buzasi\inst{3}
          }

%   \offprints{D. Stello: stello@phys.au.dk}

   \institute{Institute for Fysik og Astronomi (IFA), Aarhus Universitet,
              8000 Aarhus, Denmark
              \email{stello@physics.usyd.edu.au}
         \and
             School of Physics, University of Sydney,
             NSW 2006, Australia
         \and
             Department of Physics, US Air Force Academy,
             Colorado Springs, CO 80840, USA
             }

%   \date{Received August 15, 2005; accepted , 2005}

   \abstract{We show evidence that the red giant star $\xi\,$Hya
             has an oscillation mode lifetime, $\tau$, of about 2 days 
             significantly shorter than 
             predicted by theory ($\tau=17\,$days, \citealt{HoudekGough02}).
             If this is a general trend of red giants it would limit 
             the prospects of asteroseismology on these stars because 
             of poor coherence of the oscillations.
%             The discrepancy between theory and observation
%             implies that red giant stars can help us
%             better understand the damping and driving mechanisms of 
 %            solar-like p-modes by convection.

%   \keywords{Stars: red giants -- Stars: individual: $\xi\,$Hya -- Stars: oscillations
%               }
   }

   \maketitle
%
%________________________________________________________________

\section{Introduction}

The mode lifetime of solar-like oscillations is an important
parameter.
The interpretation of the measured oscillation frequencies (and
their scatter) relies very much on knowing the mode
lifetime, but currently we know very \mbox{little}
about how this property depends on
the stellar parameters (mass, age and chemical composition).
The theoretical estimates of mode lifetimes are based on
a simplified description of the convective environment in which
the damping and excitation of the modes takes place.
Measurements of the mode lifetime in different stars will be very
helpful for a more
thorough treatment of convection in stellar modeling.

In the following we give a short description of the method introduced by 
\citet{Stello05} to measure the large frequency separation,
$\Delta\nu_{0}$, and mode lifetime, $\tau$, of the solar-like 
oscillations in the red giant $\xi\,$Hya.
%We also quantify the ambiguity of the measured frequencies.

%__________________________________________________________________

\section{Measuring mode lifetime}\label{method}

\begin{figure}
\resizebox{\hsize}{!}{\includegraphics{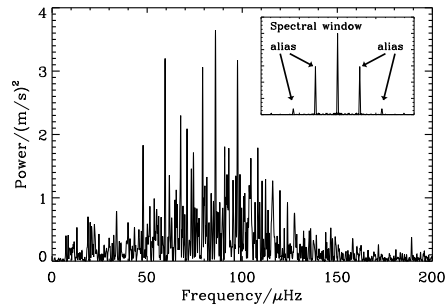}}
\caption[scaled solar spectrum]{\label{fig4}
{\footnotesize Power spectrum of $\xi\,$Hya. The frequency axis of the
spectral window in the inset is scaled to match that of the main plot.
}}
\end{figure}

We use the same data set as in \citet{Stello04}, which
is single-site radial velocity observations
covering almost 30 days.
Fig.~\ref{fig4} shows the power spectrum of the time series. 
For further details about the observations and data analysis
see \citet{Stello02,Frandsen02,Teixeira03,Stello04,Stello05}.

We assume that the mode frequencies,
$\nu_n$, of $\xi\,$Hya show a simple comb pattern 
($\nu_{n}\simeq \Delta\nu_{0}n+X_{0}$, where $\Delta\nu_{0}$ and 
$X_{0}$ are constants and $n$ is the mode order) in the power
spectrum with only radial modes.
This is supported both by theory and 
observation \citep{Dalsgaard04,Stello04}.
The finite lifetimes of the oscillations will introduce deviations
of the measured frequencies from the true mode
frequencies and hence from the comb pattern \citep{Anderson90}.
\begin{figure}
\resizebox{\hsize}{!}{\includegraphics{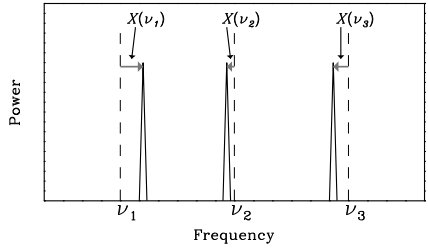}}
\caption[quality]{\label{figoffset}
{\footnotesize Schematic illustration of the frequency scatter due to a finite 
mode lifetime (see text).}}
\end{figure}
In Fig.~\ref{figoffset} we illustrate this by showing, schematically, 
a comb pattern of `true' mode frequencies (dashed lines) and the
measured frequencies (solid peaks). The deviations, indicated with
$X(\nu_{i})\equiv (\nu_{i}-X_{0})\bmod \Delta\nu_{0}$, are 
independent because each mode is excited independently.
Shorter mode lifetimes give larger deviations.

We estimate the mode lifetime from the scatter of the measured
frequencies about a regular comb pattern.
Our method does not require us to assign the mode order or degree
to the measured frequencies. They are therefore allowed 
to contain false detections from alias and noise peaks.

First, we find the comb pattern that best matches our
observed frequencies, $\nu\equiv[\nu_{1},\dots,\nu_{\mathrm{N}}]$. 
We do this by minimizing the RMS of $X(\nu)$ (called $\sigma_{X}$), with 
$\Delta\nu_{0}$ and $X_{0}$ as free parameters.
This gives us $\Delta\nu_{0}$ and $X_{0}$.

\begin{figure}
\resizebox{\hsize}{!}{\includegraphics{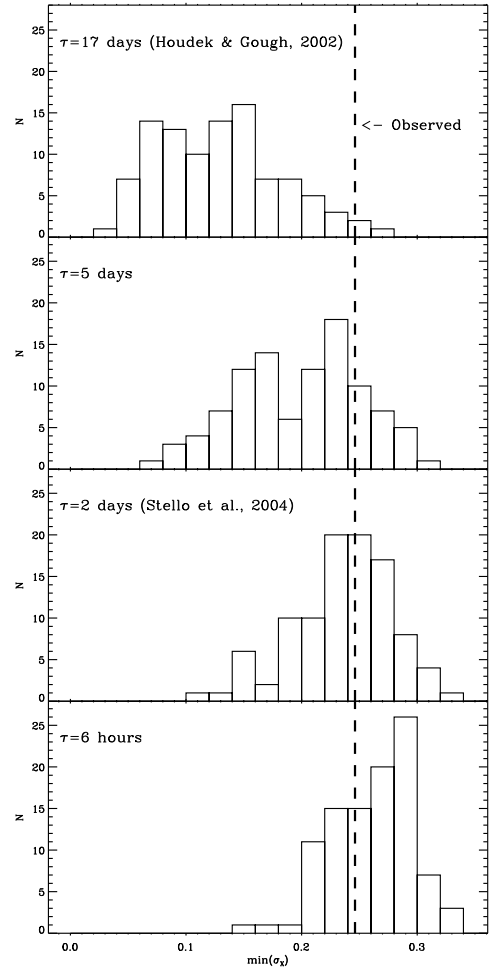}}
\caption[S(nu) distrib]{\label{fig3}
{\footnotesize Distribution of min($\sigma_{X}$)
from 100 simulations with different
mode lifetimes: 17 days, 5 days, 2 days, and 6 hours.
The dashed line indicates the value from the observations.
}}
\end{figure}

Secondly, we calibrate the minimum of $\sigma_{X}$, min($\sigma_{X}$),
against simulations with known mode lifetime.
We simulated the $\xi\,$Hya time series using the method
described in \citet{Stello04}. The oscillation mode lifetime was an
adjustable parameter, assumed to be independent of frequency, 
while the other inputs for the simulator were fixed and
chosen to reproduce the observations (see \citealt{Stello04}, Fig.~12).
For different values of the mode lifetime, we simulated 100 time series
with different random number seeds. For each we
measured 10 frequencies ($\nu_{1},\dots,\nu_{\mathrm{10}}$) using
iterative sine-wave fitting and then minimized $\sigma_{X}$.
This provided 100 values of min($\sigma_{X})$ for each mode lifetime, 
which we compare with the observations in Fig.~\ref{fig3}.
Only a few out of the 100 trials with $\tau=17\,$days (corresponding 
to the damping rate $\eta=1/(2\pi \tau)\simeq 0.1\,\mu$Hz calculated by
\citealt{HoudekGough02}) have a value for min($\sigma_{X}$) as high 
as the observations.
We find the best match with observations for mode lifetimes of
about 2 days, in good agreement with \citet{Stello04}.
However, we note that there is also a reasonable match for all mode 
lifetimes less than a day, as their distributions
all look very similar to the bottom panel.
Randomly distributed peaks also show similar distributions to
$\tau=6\,$hours. Hence, if the mode lifetime of $\xi\,$Hya is only
a fraction of a day it would definitely destroy any prospects
for asteroseismology on this star.

\begin{figure}
\resizebox{\hsize}{!}{\includegraphics[clip=true]{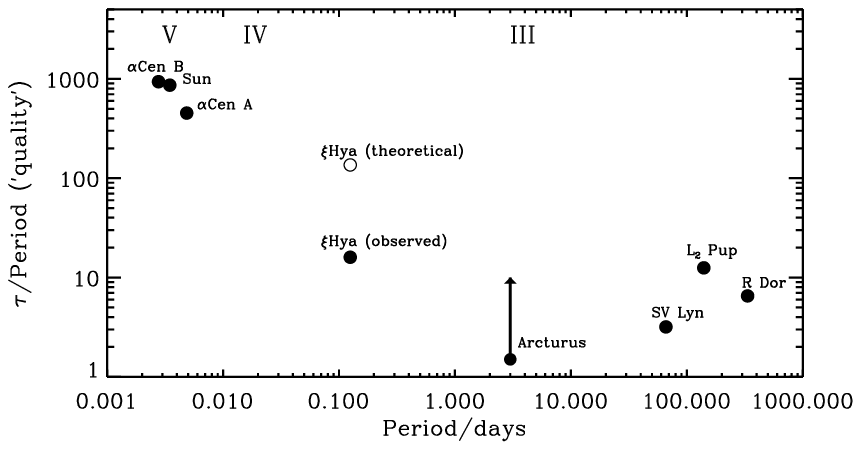}}
\caption[quality]{\label{figteas}
{\footnotesize Measured oscillation `quality' factor vs. period. 
Empty symbol shows theoretical value.
Arrow indicates lower limit. Luminosity classes are indicated.
Values from: $\alpha\,$Cen B, \citet{Kjeldsen05}; 
Sun, \citet{Chaplin97}; $\alpha\,$Cen A, \citet{Kjeldsen05}; $\xi\,$Hya 
(theoretical), \citet{HoudekGough02}; $\xi\,$Hya 
(observed), \citet{Stello05}; Arcturus, \citet{Retter03}; 
L$_2$ Pup, \citet{Bedding05a}; SV Lyn and R Dor, \citet{Dind04}.}}
\end{figure}
In Fig.~\ref{figteas} we plot the ratios between the measured 
mode lifetime and period (the oscillation `quality' factor) 
as a function of period for selected stars. 
Roughly speaking, the `quality' factor is 
the number of cycles over which the oscillation is coherent,
%the number of oscillation cycles with constant phase, 
and the higher this number, 
the better we can determine the frequency. 
The observed point for $\xi\,$Hya indicates there is a steep decline 
in the quality factors for stars above the main 
sequence in the Hertzsprung-Russell diagram. If this is a general 
trend  it will limit the extent by which 
we can use asteroseismology on these more evolved stars.

%The discrepancy between theory and observation in the mode lifetime of 
%$\xi\,$Hya implies that red giants could be used
%to better understand the mechanisms of the driving and damping of
%solar-like p-modes in a convective environment.

%__________________________________________________________________

\section{Frequency analysis}

\begin{figure}
\resizebox{\hsize}{!}{\includegraphics{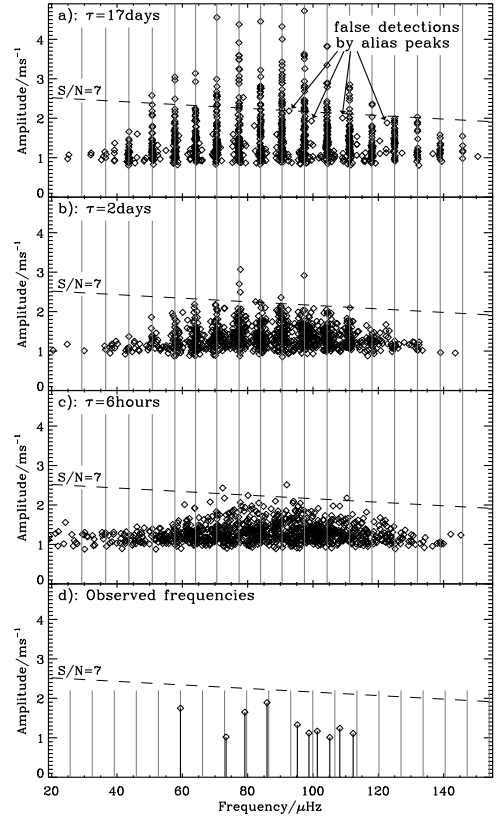}}
\caption[unambiguity]{\label{fig7}
{\footnotesize Each panel shows 1000 measured frequencies and their amplitudes
(10 from each of 100 simulations).
Mode lifetimes are indicated in each panel.
Dashed lines indicate 7 times the
input noise, and the solid grey lines are the input frequencies.
Panel (d): Observed frequencies and the best matching comb pattern 
(grey lines).}}
\end{figure}
We investigated the reliability of the observed frequencies
using the simulations
described in Sect.~\ref{method}. The measured frequencies,
together with the input frequencies and noise level are plotted 
in Fig.~\ref{fig7}.
Apart from the broadening of the mode frequencies due to damping,
we also see false detections of alias peaks. This is most easily
seen in panel (a) where the damping is less 
(a few examples are indicated).
Our test shows that frequencies are not unambiguous
if S/N$\lesssim$7--8, even for mode lifetimes of 17 days.
For short mode lifetimes it looks rather hopeless to measure
the individual mode frequencies with useful accuracy.
The observed frequencies are all with S/N$<$6.3 and hence cannot be
claimed to be unambiguous.
The lack of a clear comb pattern similar to Fig.~\ref{fig7} (top panel)
in the observed frequencies (bottom panel) also supports a short mode 
lifetime.

%__________________________________________________________________

%\section{Discussion and future prospects}\label{discussion}

\begin{figure}
\resizebox{\hsize}{!}{\includegraphics{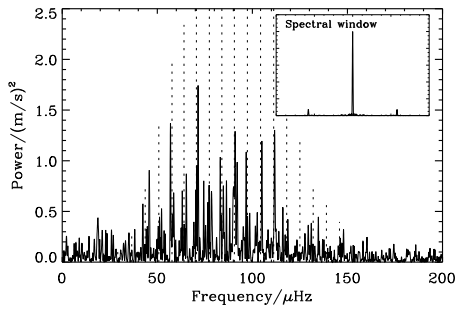}}
\caption[sim multisite]{\label{fig8}
{\footnotesize Power spectrum of simulated two-site time series of 
$\xi\,$Hya ($\tau=2\,$days).
Dotted lines indicate the input frequencies.
 }}
\end{figure}

A difficulty in using the current observations of $\xi\,$Hya for 
asteroseismology
arises from the severe crowding in the power spectrum. The crowding  
comes partly from the single-site spectral window,
which emphasizes the importance of using more continuous
data sets.
To illustrate this, we made a power spectrum of a simulated 
two-site time series shown in Fig.~\ref{fig8}.
Each mode profile is seen much more clearly, though slightly
blended due to the short mode lifetime.
Obviously, more can be obtained from such a spectrum than from
our present data set (Fig.~\ref{fig4}), but a thorough
analysis of similar simulations should be done to determine the prospects
for doing asteroseismology on $\xi\,$Hya.

%__________________________________________________________________

\section{Conclusions}\label{conclusion}

We find that the most likely mode lifetime of $\xi\,$Hya is about 2 days, and
we show that the theoretical prediction of 17 days \citep{HoudekGough02}
is unlikely to be the true value.

Due to the high level of crowding in
the power spectrum, the signature from the p-modes
is too weak to determine the
large separation to very high accuracy.
However, our measurement supports the separation
of $6.8\,\mu$Hz found by \citet{Frandsen02}.

We conclude that the only quantities we can reliably obtain from
the power spectrum of $\xi\,$Hya are the mode amplitude, mean mode
lifetime, and the average large frequency separation. 

Our simulations show that
none of the measured frequencies from
the $\xi\,$Hya data set \citep{Frandsen02} can be regarded as unambiguous.
Only in the case of a greatly improved window function could
it be possible to detect frequencies unambiguously.

\begin{acknowledgements}
This work was supported in part by the Australian Research Council.
\end{acknowledgements}

%__________________________________________________________________

\bibliography{bib_complete}

\end{document}